%% file: iclr2025_conference.tex
\documentclass{article} % For LaTeX2e
\usepackage{iclr2025_conference,times}

\input{math_commands.tex}

\usepackage{hyperref}
\usepackage{url}
\usepackage{float}
\usepackage{microtype}
\usepackage{graphicx} % Required for inserting images
\usepackage{amsmath}
\usepackage[utf8]{inputenc} % allow utf-8 input
\usepackage[T1]{fontenc}    % use 8-bit T1 fonts
\usepackage{hyperref}       % hyperlinks
\usepackage{url}            % simple URL typesetting
\usepackage{booktabs}       % professional-quality tables
\usepackage{amsfonts}       % blackboard math symbols
\usepackage{nicefrac}       % compact symbols for 1/2, etc.
\usepackage{microtype}      % microtypography
\usepackage{xcolor}         % colors
\usepackage{xspace}
\usepackage{amsmath}
\usepackage{amssymb}
\usepackage{booktabs}
\usepackage{makecell}
\usepackage{colortbl}
\usepackage{graphicx}    
\usepackage{multirow}
\usepackage{interval}
\intervalconfig{soft open fences}
\usepackage{algorithm, algpseudocode}

\usepackage{subfigure}
\usepackage{float}
\usepackage{enumitem}
\usepackage{tabularx}
\usepackage{placeins}
\usepackage{color}
\usepackage{tcolorbox}
\usepackage{colortbl} 
\usepackage{listings}
\usepackage{makecell}
\usepackage{siunitx}

\author{%
  Zitian Gao \\
  The University of Sydney\\
  \texttt{zgao3186@uni.sydney.edu.au} \\
  \And
  Yihao Xiao \\
  Shanghai University of Finance and Economics \\
  \texttt{2023121716@stu.sufe.edu.cn} \\
}

\title{Enhancing Startup Success Predictions in Venture Capital: A GraphRAG Augmented Multivariate Time Series Method}

\iclrfinalcopy % Uncomment for camera-ready version, but NOT for submission.

\begin{document}

\maketitle

\begin{abstract}
In the Venture Capital (VC) industry, predicting the success of startups is challenging due to limited financial data and the need for subjective revenue forecasts. Previous methods based on time series analysis often fall short as they fail to incorporate crucial inter-company relationships such as competition and collaboration. To fill the gap, this paper aims to introduce a novel approach using GraphRAG augmented time series model. With GraphRAG, time series predictive methods are enhanced by integrating these vital relationships into the analysis framework, allowing for a more dynamic understanding of the startup ecosystem in venture capital. Our experimental results demonstrate that our model significantly outperforms previous models in startup success predictions.
\end{abstract}

\input{tex/intro}
\input{tex/related_work}
\input{tex/method}
\input{tex/experiment}

\input{tex/conclusion}

\bibliography{iclr2025_conference}
\bibliographystyle{iclr2025_conference}
\end{document}

%% file: math_commands.tex
\usepackage{amsmath,amsfonts,bm}

\def\eqref#1{equation~\ref{#1}}
\def\1{\bm{1}}

\DeclareMathAlphabet{\mathsfit}{\encodingdefault}{\sfdefault}{m}{sl}
\SetMathAlphabet{\mathsfit}{bold}{\encodingdefault}{\sfdefault}{bx}{n}

%% file: tex/intro.tex
\section{Introduction}
Startups typically represent newly established business models associated with disruptive innovation and high scalability. They are often viewed as powerful engines for economic and social development~\cite{loukas-etal-2023-using}. With the ongoing economic and social development and the continuous emergence of new technologies, the number of startups has surged, making the prediction of new technologies or business models highly uncertain. Venture capital (VC) evaluation involves assessing investment opportunities in startups and early-stage companies, requiring effective decision-making and data analysis. Valuation is crucial for VCs when deciding to invest in a startup, yet nearly 50\% of early-stage VCs make intuitive investment decisions using qualitative methods like gut feelings \cite{GOMPERS2020169}. To avoid relying solely on human expertise and intuition, investors often adopt data-driven approaches to predict the success probability of startups~\cite{COREA2021100062,hunter2018pickingwinnersdatadriven}.

Previous work has primarily focused on simple deep learning and machine learning methods, which often rely on limited financial data or single time series analyses~\cite{wang2024automatedstartupevaluationpipeline,lewis2021retrievalaugmentedgenerationknowledgeintensivenlp}. Specifically, most models focus on using historical financial indicators, market trends, or other single-dimensional data to predict the future performance of startups. While these models can capture the growth potential of startups in some cases, they often overlook critical factors within the broader ecosystem in which startups operate, especially the competitive and cooperative relationships between companies. Startups typically operate in environments with limited resources, and their interactions with one another play a significant role in their success or failure. Traditional methods have failed to fully leverage this relational data, leading to poor performance in predicting startup success, particularly in cases with sparse data and low signal strength. These models struggle to accurately identify early success signals under such conditions.

To address these challenges, we propose a novel approach that combines GraphRAG technology with multivariate time series analysis to improve the prediction of startup success. The core of GraphRAG technology lies in its ability to effectively integrate information from various sources, particularly through knowledge graphs and retrieval-augmented generation models~\cite{edge2024localglobalgraphrag}, to deeply explore the complex network of relationships among startups. Knowledge graphs can showcase collaborations, competition, supply chain relationships between companies, and other critical information essential for understanding a company's positioning within its industry ecosystem. By combining this relational information with multivariate time series data, we can build more comprehensive and sophisticated prediction models, significantly improving prediction accuracy and performance on companies with sparse data. Specifically, traditional time series analysis methods often struggle to provide sufficient information support for startups with sparse data and limited samples, leading to instability and low accuracy in prediction results. However, GraphRAG technology enhances model performance by integrating multi-source data, particularly incorporating relationship information among companies into the prediction models, allowing for supplementation and enhancement even in sparse data conditions.

\begin{figure*}[t!]
\centering
\includegraphics[width=\textwidth]{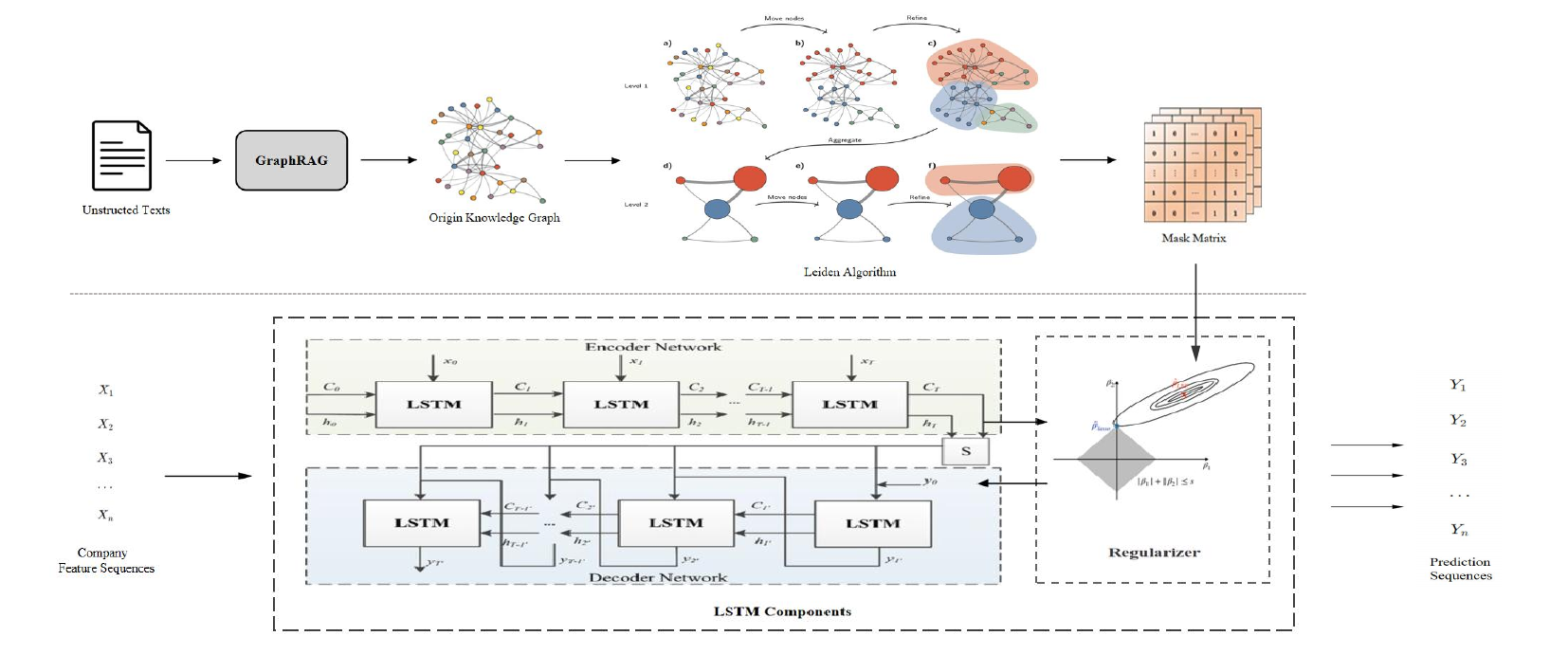}
\caption{Our proposed method framework overview: The upper part of the framework represents the process of extracting a knowledge graph from structured text using GraphRAG, followed by transforming it into a mask matrix using the Leiden algorithm. The lower part represents our multivariate sequence-to-sequence LSTM model \cite{Mou}.}
\label{fig:approach}
\end{figure*} 

Startups have various potential development goals, with the most desirable being successful acquisition by a large company or going public. While many studies and VCs focus on predicting whether a company will successfully go public as a binary classification task~\cite{Cao_2022,wang2024automatedstartupevaluationpipeline}, we attempt to predict what will happen to the company next. This approach aims to provide VCs with opportunities not only to consider high-risk, high-reward companies but also to build lower-risk investment portfolios and determine when to exit~\cite{exit1,exit2}.

This paper utilizes two publicly available Chinese news text datasets~\cite{he2023wanjuancomprehensivemultimodaldataset,lu2023bbtfincomprehensiveconstructionchinese}, selecting data and unstructured news text from companies listed on the A-share market in China from 2013 to 2024. We extracted news reports related to A-share listed companies from these datasets, combined with these companies' financial data, market performance, and other relevant time series data to construct a comprehensive dataset. Various quantitative and qualitative factors influence the success of startups. In constructing the dataset, we not only collected news reports related to A-share listed companies for the specified years but also conducted an in-depth analysis of the various quantitative and qualitative factors mentioned in these reports. These factors include but are not limited to changes in company executives, industry dynamics, market competition, intellectual property (such as patents, trademarks, etc.). These news reports were transformed into quantifiable indicators using large models for Named Entity Recognition (NER) and structured output, allowing for effective integration with financial data, market performance, and other structured data. By integrating these quantitative and qualitative factors, we strive to build a predictive model that comprehensively reflects the potential of startups.

In contrast to another common direction in financial time series prediction—stock trend forecasting—fine-grained, structured time series data used for predicting stock trends is readily available~\cite{stock1,stock2,stock3}. By comparison, predicting the success of startups in the venture capital (VC) field requires forecasting the success of unlisted startups with minimal available data, often relying on hard-to-access unstructured text such as team backgrounds, patents, and product reviews.

Stock trend forecasting typically utilizes fine-grained data at the second or minute level, whereas startup success prediction in the VC domain often relies on coarse-grained data at the quarterly or annual level. Additionally, relationships such as collaboration, competition, and investment activity play a more significant role in the VC industry compared to stock forecasting in quantitative finance. These distinctions pose unique challenges for predicting startup success in the VC field~\cite{coop1,coop2}.

To address these issues, we proposed our GraphRAG
augmented multivariate time series method. \textbf{Our main contributions are as follows:}

\begin{enumerate}
    \item We propose a GraphRAG-enhanced multivariate Seq2Seq time series analysis method to predict startup success in the context of venture capital (VC).
    \item We review previous related work and identify the lack of crucial relationship information, such as collaboration and competition, in existing methods. We propose using GraphRAG to extract these relationships from unstructured text data and integrate them into the multivariate time series analysis framework.
    \item We introduce our GraphRAG-based data pipeline, which extracts structured information from extensive news text to address the venture capital industry's data scarcity for multivariate time series analysis. We then present a novel method to integrate key knowledge graph information, significantly improving startup success prediction performance.
\end{enumerate}

% The remainder of this paper is organized as follows: Related Work section reviews the related literature, focusing on startup success prediction, venture capital analysis, and the application of knowledge graphs and multivariate time series models. We analyze the limitations of existing prediction models and discuss recent research progress in these fields and their potential in venture capital. Proposed method section details our methodology. Experiment introduces the data collection and processing process, and discusses the experimental setup and results of model comparisons. Finally, the last sections summarize our contributions and propose limitations and future research directions.

%% file: tex/related_work.tex
\section{Related Work}

\subsection{Startup Success Prediction\\}
Predicting the success of startups has garnered increasing attention from scholars, especially in the venture capital field. Existing research primarily focuses on developing methods to represent company features and relationships and on building predictive models using machine learning and deep learning techniques. For example, \cite{Cao_2024} constructed CompanyKG to represent and learn various features and relationships between companies, aiming to capture the impact of these features on a company's success. \cite{ZBIKOWSKI2021102555} used Crunchbase data to create an XGBoost-based model for predicting the likelihood of a company's success. \cite{KIM2020113229} used a doc2vec deep learning method to extract feature vectors from startup profiles and patent abstracts, applying them to filter promising startups in certain fields.In addition, \cite{ROSS202194} introduced the CapitalVX framework, which leverages general business information of startups to predict optimal entry and exit timings, providing valuable insights into market dynamics and investment strategies.  Although these approaches have made some progress, they still face challenges in handling complex nonlinear relationships and long-term predictions. \cite{lyu2021graphneuralnetworkbased} designed an incremental representation learning mechanism and a sequential learning model based on graph neural networks to predict the success of startups. \cite{Wang2020} performed preprocessing and exploratory analysis on Kickstarter data and then introduced deep learning algorithms like multilayer perceptrons for predicting crowdfunding outcomes. To address long-term revenue prediction, \cite{Cao_2022} proposed the Simulation of Information Revenue Extrapolation (SiRE) algorithm to generate fine-grained long-term revenue forecasts.Furthermore, some studies, such as \cite{Leogrande2021}, have explored the use of unstructured data, including news articles and social media content, for predicting startup success, highlighting the potential of diverse data sources in this domain.

These studies provide valuable insights into startup success prediction and demonstrate the potential of deep learning in this domain. However, these models typically require large amounts of training data and may perform poorly in scenarios with sparse or noisy data. Existing methods also have limitations in integrating multi-dimensional features and relationship information, making it challenging to fully leverage the latent connections and temporal relationships within complex company ecosystems. Against this backdrop, our research seeks to overcome the limitations of traditional methods by combining GraphRAG with multivariate time series analysis to provide a more comprehensive and accurate solution for predicting startup success.

\subsection{Retrieval-Augmented Generation In Finance\\}
In the finance domain, Retrieval-Augmented Generation \cite{lewis2021retrievalaugmentedgenerationknowledgeintensivenlp} has become a significant method, particularly for answering questions based on large datasets. RAG methods extract information from retrievable text regions, providing a solid foundation for generation tasks. For instance, \cite{wang2024automatedstartupevaluationpipeline} developed a startup success prediction framework (SSFF) that incorporates RAG as an external knowledge module to gain a comprehensive understanding of market landscapes. However, in more complex scenarios, such as handling multi-dimensional data and intricate relationships, the capabilities of traditional RAG may be limited.

To address this issue, GraphRAG has been introduced as a structured, hierarchical retrieval-augmented generation method. It not only relies on semantic search of plain text fragments but also enhances the model's reasoning ability by constructing and utilizing knowledge graphs \cite{edge2024localglobalgraphrag}. The GraphRAG process involves extracting information from all kinds of unstructured text data (e.g., PDFs, CSVs), building knowledge graphs, generating hierarchical summaries of data, and using the Leiden algorithm to detect industry communities within the knowledge graph \cite{Traag_2019}. This process aids in intuitively understanding the collaborative and competitive relationships between companies when dealing with complex, multi-dimensional data.

GraphRAG has shown significant advantages in reasoning about complex information, especially in handling relationship networks and time series data between startups. Compared to traditional RAG, GraphRAG can better connect different pieces of information, generating new synthesized insights through shared attributes and providing more accurate and in-depth predictive results. This approach is particularly effective when dealing with previously unseen proprietary datasets, such as a company's proprietary research or business documents. Traditional RAG struggles to fully comprehend and summarize large datasets, whereas GraphRAG, by automatically creating knowledge graphs and combining community summaries with graph machine learning outputs, significantly improves task performance.

In our research, GraphRAG is applied to multivariate time series analysis for startup success prediction. As a key component, it is used to extract relationship information, such as collaboration and competition, from unstructured text data and integrate it into our predictive model. This approach not only enhances the model's reasoning capabilities but also improves the accuracy and effectiveness of predictions by better capturing the complex relationships between startups.

%% file: tex/method.tex
\section{Methodology}
Our approach consists of two stages: First, we input a large amount of unstructured news text into GraphRAG, where during the clustering and retrieval process, global statistics are dynamically updated to accurately cluster relationships between entities. This results in a directed knowledge graph of companies, with edges representing both direct and indirect relationships between companies. In the second stage, inspired by \cite{ibrahim2022knowledge} and \cite{Barigozzi2019}, we transform the knowledge graph into a mask matrix using the Leiden algorithm \cite{Traag_2019}. This matrix is then used as a regularizer in the multivariate time series model, combined with relevant covariates. This method effectively enhances the generalization ability of models on scarce company data and significantly improves prediction performance, as illustrated in Figure 1. The specific algorithm details are provided in Algorithms 1 and 2.

\subsection{Sequence-to-Sequence Representation\\}
We define the startup success prediction as a sequence-to-sequence prediction task, which fundamentally differs from prior approaches that typically focus on binary classification tasks, such as predicting whether a company will go public or not. Unlike these methods, which often use deep learning models to directly learn a binary outcome, our approach captures a more nuanced and comprehensive understanding of a startup's long-term performance by predicting a sequence of outcomes.

Given a company \( X \), the input sequence \( X_n \) comprises a rich set of feature sequences spanning from 5 to 10 years prior to the company's IPO. These features include a variety of dimensions such as patents, trademarks, legal cases, funding history, executive changes, and other key factors that reflect the company's operational environment and strategic decisions. These dimensions are not simply individual indicators but are treated as multivariate representations that collectively capture the company's growth trajectory, strategic shifts, and adaptability to market conditions.

The output sequence \( Y_n \) represents the company's post-IPO performance over several quarters, specifically in terms of the price-to-book ratio (P/B ratio). The P/B ratio, defined as the ratio of the company's stock price (\textit{Price}) to its book value (\textit{Book Value}), serves as a crucial metric in venture capital. Unlike the price-to-earnings ratio (P/E ratio), which can be volatile in the short term due to fluctuations in earnings, the P/B ratio offers a more stable measure of a company's intrinsic value and long-term sustainability \cite{Nissim}. By predicting the P/B ratio over multiple quarters, our model provides a detailed projection of a company's ongoing financial health and shareholder profitability.

This sequence-to-sequence approach allows us to move beyond the limitations of binary classification, enabling a deeper analysis of a company's potential success trajectory. By focusing on continuous, multivariate output sequences rather than a single binary outcome, our method offers a richer, more detailed predictive framework that is better suited to the complex, dynamic nature of startup ecosystems. This approach also allows us to capture the temporal evolution of key performance indicators, providing venture capitalists with a more informed basis for their investment decisions.

% \begin{figure}
% \centering
% \includegraphics[width=0.47\textwidth]{fig/fig2.pdf}
% \caption{The impact of different mask matrix regularization strengths \( \lambda \) on the model's prediction performance (R-squared).}
% \label{fig:2}
% \end{figure}

\begin{figure*}[t!]
\centering
\includegraphics[width=0.62\textwidth]{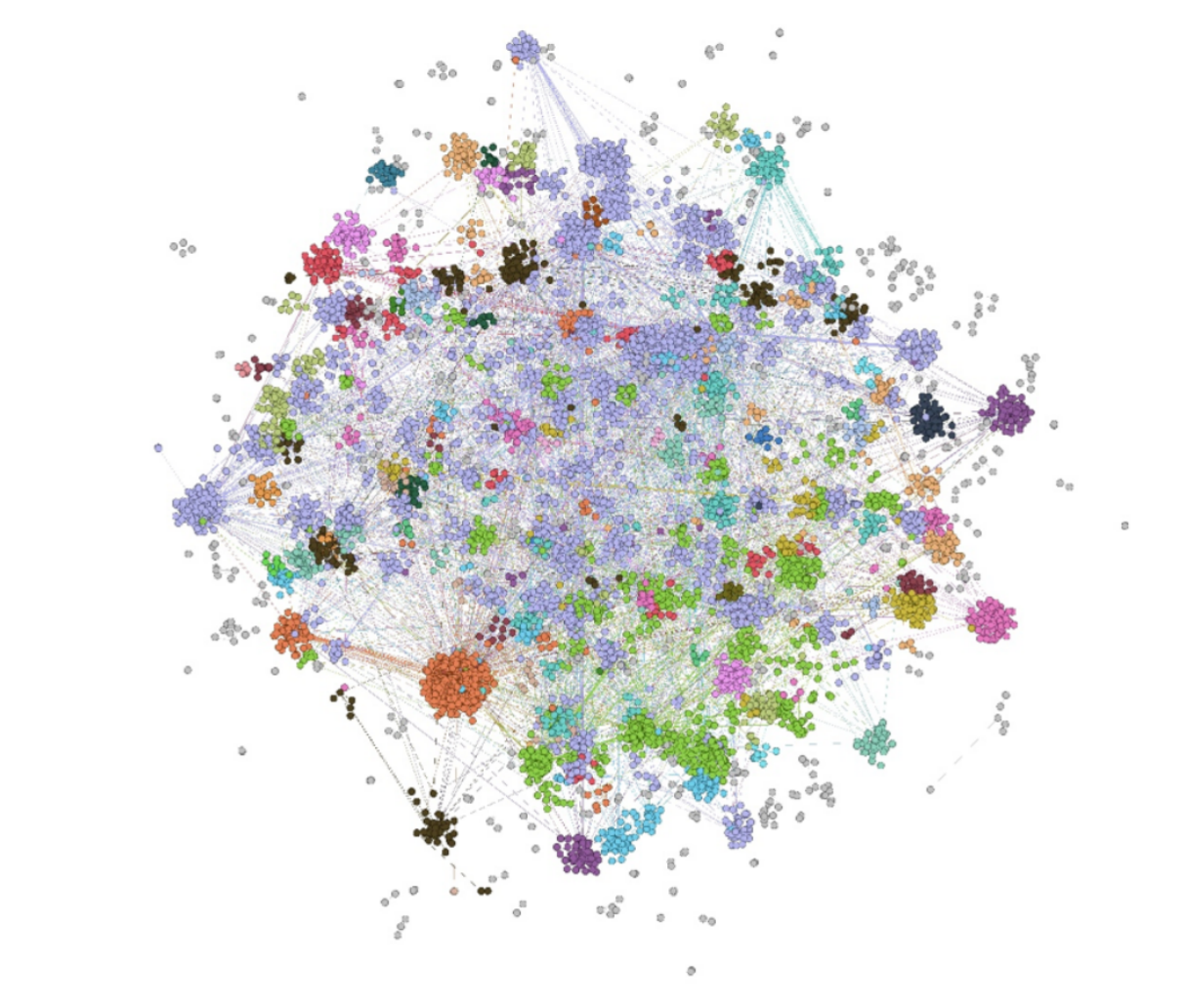}
\caption{The knowledge graph generated using GraphRAG and the Leiden algorithm, illustrated using OpenORD \cite{openord}, different colors represent different industries (community clusters).}
\label{fig:3}
\end{figure*}

\subsection{Graph Clustering\\}
Let $G(V, E)$ be an undirected graph where \( n := |V| \) represents the number of vertices, and \( m := |E| \) represents the number of edges. We define \( C = \{C_1, \dots, C_k\} \) as a partition of the vertex set \( V \). The set \( C \) is referred to as a clustering of \( G \), with each \( C_i \) being non-empty. A clustering \( C \) is termed trivial if \( k = 1 \) or \( k = n \). The collection of all possible clusterings of \( G \) is denoted by \( \mathcal{A}(G) \). In what follows, a cluster \( C_i \) is frequently associated with the induced subgraph of \( G \), denoted as \( G[C_i] := (C_i, E(C_i)) \), where
\[ E(C_i) := \{\{v, w\} \in E : v, w \in C_i\}. \]
Thus, we define \( E(C) := \bigcup_{i=1}^k E(C_i) \) as the collection of intracluster edges, while the complement set \( E \setminus E(C) \) represents the intercluster edges. The total number of intracluster edges is denoted by \( m(C) \) with the total number of intercluster edges by \( \overline{m}(C) \). Additionally, the set of edges that connect a node in \( C_i \) to a node in \( C_j \) is denoted by \( E(C_i, C_j) \).

In complex networks, nodes tend to group together, forming relatively dense clusters commonly referred to as communities. Modularity is a measure used to assess the quality of communities identified by heuristic-based community detection algorithms. The formulation of modularity from \cite{4358966} is as follows:

\begin{scriptsize}
\begin{align*}
{q}({C}) := & \sum_{C \in {C}}\left[{{|E({C})|}\over{m}} - 
\left({{|E({C})| + \sum_{C^{\prime} \in {C}}|E(C, C^{\prime})|}\over{2m}}\right)^{2}\right].
\end{align*}
\end{scriptsize}

The Leiden algorithm originally introduced by \cite{Traag_2019} is a multi-level approach to community detection aiming to maximizing the modularity gain. This algorithm involves three stages. The first stage is known as the local-moving phase, involves each vertex \(i\) optimizing its community assignment by greedily selecting to join the community of one of its neighbors denoted as \(j \in J_i\). The next stage is the refinement phase, where vertices within each community further update their community assignments, starting from singleton partitions. Unlike the local-moving phase, these updates are not purely greedy; instead, vertices may move to any community within their scope where the modularity gain is positive, with the likelihood of moving to a neighboring community proportional to the modularity gain of that move. Finally, the aggregation phase combines all vertices within each refined partition into super-vertices, with an initial community assignment inherited from the local-moving phase \cite{Traag_2019}.

The algorithm introduces a refinement phase subsequent to the local-moving phase, wherein vertices within each community undergo constrained merges to other sub-communities within their community bounds (obtained from the local-moving phase), starting from a singleton sub-community. This is performed in a randomized manner, with the probability of joining a neighboring sub-community within its community bound being proportional to the delta-modularity of the move. This facilitates the identification of sub-communities within those obtained from the local-moving phase. Once communities have converged, it is guaranteed that all vertices are optimally assigned, and all communities are subset optimal \cite{Traag_2019}.

Since startup success prediction in venture capital often uses a quarter as the time unit, the performance differences between time series models are usually minimal. Instead, we focus more on how to leverage information from the GraphRAG-produced knowledge graph to enhance the performance of time series models.

\subsection{Mask Matrix Regularization\\}
Recall the origin Lasso \cite{51791361-8fe2-38d5-959f-ae8d048b490d} and adaptive Lasso 
 \cite{doi:10.1198/016214506000000735} formulations:

\textbf{Lasso:}
\[
\hat{\beta}^\mathcal{L}_n = \arg\min_{\beta} \left\{ \frac{1}{n} \|y - X\beta\|_2^2 + \frac{\lambda_n}{n} \sum_{j} |\beta_j| \right\}
\]
where $\|y - X\beta\|_2^2$ denotes the residual sum of squares, and $\sum_{j} |\beta_j|$ is the $\ell_1$-norm penalty term.\\

\textbf{Adaptive Lasso:}
\begin{small}
\[
\hat{\beta}^A_n = \arg\min_{\beta} \left\{ \frac{1}{n} \|y - X\beta\|_2^2 + \frac{\lambda_n}{n} \sum_{j} \frac{1}{|\hat{\beta}_j|^\gamma} |\beta_j| \right\}
\]
\end{small}
where $\mathbf{r}$ is the element-wise reciprocal of the pre-estimator of $\rho$. After we extract the sparse connectivity structures $\mathcal{E}$ from $G$ using Leiden algorithm, we define the mask matrices formulation from \cite{ibrahim2022knowledge} as follows:
\[
M_{ij} =  1, \text{if } (i,j) \in \mathcal{E}; M_{ij} = \infty, \text{otherwise}.
\]
These masking matrices \( M \) are used to apply modified regularization penalties to \( \rho \). As the masking entries increase, the corresponding penalties applied to the entries in \( \rho \) also increase. Larger entries effectively zero out the corresponding elements in \( \rho \), thereby reducing model complexity. With these additional masking weights, we use the newly redefined masked matrix regularizer from \cite{ibrahim2022knowledge} as follows:
\[
\Omega_\rho(\rho) = \lambda_\rho \|\mathbf{M} \odot \mathbf{r} \odot \rho\|_1.
\]
where $\mathbf{r}$ is the element-wise reciprocal of the prior knowledge/pre-estimator of $\rho$.\\

Based on the ablation study results in Section~\ref{ablation}, our new masked regularizer were found to be effective in improving model performance by enforcing sparsity and preserving important structural information in the graph. This regularizer not only enhanced the interpretability of the learned representations but also demonstrated robustness against overfitting, particularly in high-dimensional and complex datasets.

%% file: tex/experiment.tex
\begin{table*}[t]
\label{table-1}
\begin{center}
\renewcommand{\arraystretch}{1.2}
\setlength{\tabcolsep}{18pt}
\resizebox{1.0\textwidth}{!}{  
\small
\begin{tabular}{l|cccccc}
\hline
\multicolumn{1}{l|}{\textbf{Model}}
&\multicolumn{1}{c}{MSE} 
&\multicolumn{1}{c}{MAE}
&\multicolumn{1}{c}{RMSE} 
&\multicolumn{1}{c}{p-value} 
&\multicolumn{1}{c}{R-squared}\\
\hline
GRU & 0.8311 & 0.1264 & 0.9116 & 1.35e-18 & 0.3158\\
RNN & 0.9034 & 0.1190 & 0.9504 & 4.50e-24 & 0.2860\\
BERT & 0.8054 & 0.1083 & 0.9467 & 1.22e-25 & 0.3095\\
Transformer & 0.7921 & 0.1021 & 0.9514 & 2.16e-24 & 0.3275\\
LSTM & 0.7687 & 0.1053 & 0.9736 & 1.19e-29 & 0.3354\\
\hline
\textbf{Ours} & \textbf{0.6021} & \textbf{0.0832} & \textbf{0.7923} & \textbf{2.19e-44} & \textbf{0.4075}\\
\hline
\end{tabular}
}
\caption{P/B ratio prediction performance on the benchmark datasets.}
\end{center}
\end{table*}

\section{Experiment}
\subsection{Datasets\\}
Our dataset is composed of two parts: BBT-FinCorpus \cite{lu2023bbtfincomprehensiveconstructionchinese} and Wanjuan1.0 \cite{he2023wanjuancomprehensivemultimodaldataset}. BBT-FinCorpus is a large-scale Chinese financial corpus that includes approximately 300GB of raw text data gathered from four distinct sources: (1) financial, political, and economic news published by mainstream media platforms over the past 20 years; (2) all announcements and financial reports from publicly listed companies; and (3) millions of research reports from financial institutes and consulting agencies. These sources provide a comprehensive dataset that covers a wide range of financial and economic topics, offering rich background information and multidimensional data for our research. The second part of the dataset comes from Wanjuan1.0, which includes 7 million news articles covering various fields across the Chinese internet. The combination of these two datasets gives us a broad and in-depth perspective, enabling a multi-faceted analysis and prediction of startup success.

\subsection{Data Processing Pipeline\\}
However, processing such a massive dataset using the GPT-4o mini API \cite{gpt4oMINI} is quite expensive with costs amounting to \$5.00 per 1 million tokens for input and \$15.00 per 1 million tokens for output. To optimize processing efficiency and control costs, we implemented a series of data filtering measures. First, we filtered the data based on length, removing any text longer than 10,000 characters. This significantly reduced the data volume, making subsequent processing more manageable. Next, we focused on filtering data for over 300 companies that were listed on the Chinese A-share market between 2023 and 2024. We created a list of these company names and used it to search for matching fields in the news articles, removing any data that was entirely unrelated to these companies, thereby further streamlining the dataset.

To ensure the relevance and quality of the data, we employed the Deepseek-chat \cite{deepseekai2024deepseekllmscalingopensource} API’s JSON formatting output function to assign a relevance score ranging from 0 to 5 for each news article based on its importance relative to the company names. We chose to retain only those articles with higher relevance scores and removed those with a score of 0. This meticulous filtering process resulted in a refined dataset of 40,000 news articles, with an average length of 3,000 tokens per article.

Due to the substantial token consumption required by GraphRAG, the indexing process for this filtered news dataset consumed approximately 1.5 billion tokens, costing around \$270. Upon completing the indexing process, we generated multiple directed knowledge graphs that encompass entities, communities, datasets, and more, along with a significant number of community reports. The creation of these knowledge graphs not only provided structured data support for our research but also generated detailed community reports, further enriching our analytical perspectives. These knowledge graphs and the accompanying community reports have enabled us to better understand and analyze the complex relationships among startups and enhancing the accuracy and effectiveness of our prediction models.

\subsection{Experimental Setup}
In our experimental setup, we emphasize the input and output sequences discussed earlier. Specifically, the input sequence \(X_n\) includes various feature sequences from 5 to 10 years prior to a company's IPO, covering multiple dimensions such as patents, trademarks, legal cases, and funding history. These features serve as multivariate representations, capturing the company's growth trajectory across different aspects. The output sequence \(Y_n\) represents the price-to-book ratio (P/B ratio) over several quarters following the IPO, a key metric in assessing the continuous profitability of the company's shareholders, and is more stable than the price-to-earnings ratio (P/E ratio).

Our experimental setup considers the specific context of the Chinese A-share market, utilizing unstructured textual data and publicly available structured data such as intellectual property information from 2013 to 2024. We designate the period from 2020 to 2021 as the training period, 2022 as the hyperparameter validation period, and 2023 to 2024 as the final evaluation period. This temporal segmentation helps ensure the model's generalizability across different market conditions.

We evaluate the predictive performance of our models using several metrics, including Mean Squared Error (MSE), Root Mean Squared Error (RMSE), Mean Absolute Error (MAE), p-value, and R-squared. For the baseline model, we perform a grid search to identify the optimal hyperparameter configuration, using a standard Lasso regularizer.

Similarly, our proposed model also utilizes grid search to select the optimal hyperparameter configuration. And the regularizer is enhanced by our proposed mask matrix modified Lasso regularizer. This approach allows us to fully leverage inter-company relationship information, improving the model's predictive capabilities, especially under sparse data conditions.

% \subsection{Main Results}
As shown in Table 1, our proposed GraphRAG augmented sequence-to-sequence LSTM method outperforms all baseline methods on this task. Specifically, we observed an approximately 16\% improvement in R-squared compared to the best-performing model among the baselines, which fully demonstrates the effectiveness of our approach. This result highlights our model's superior ability to capture complex inter-company relationships and handle sparse data, leading to significantly better prediction accuracy. Moreover, our model exhibited robustness and reliability across various data sparsity conditions, further validating the potential of GraphRAG technology in the venture capital domain.

\subsection{Ablation Study}
\label{ablation}
We next demonstrate through two sets of ablation experiments that the information produced by the GraphRAG-generated knowledge graph effectively enhances the time series prediction of startup success. In the first set of experiments, we adjust the regularization strength of the Mask Matrix Regularization by gradually increasing the size of the regularization term, thereby controlling the amount of knowledge graph information introduced into the time series model. In the second set of experiments, we randomly remove edges from the knowledge graph, representing relationships such as collaboration, competition, and investment between entities, to show that incorporating these relationships significantly improves the performance of time series prediction for startup success.

\begin{table}[ht]
\centering
\label{tab:regularization}
\begin{tabular}{cccc}
\hline\hline
$\lambda$ & MSE & MAE & R-squared \\
\hline
0.1       & 0.6213 & 0.0901 & 0.3552 \\
0.5       & 0.6135 & 0.0876 & 0.3948 \\
1.0       & 0.6054 & 0.0854 & 0.4019 \\
5.0       & 0.6027 & 0.0840 & 0.4061 \\
10.0      & \textbf{0.6021} & \textbf{0.0832} & \textbf{0.4075} \\
\hline\hline
\end{tabular}
\caption{Effect of Regularization Strength $\lambda$ (optimal 10) on MSE, MAE, and R-squared}
\end{table}

\begin{table}[ht]
\centering
\label{tab:unremoved_nodes}
\begin{tabular}{cccc}
\hline\hline
$\alpha$ & MSE & MAE & R-squared \\
\hline
0.2       & 0.6589 & 0.0987 & 0.3402 \\
0.4       & 0.6348 & 0.0934 & 0.3625 \\
0.6       & 0.6175 & 0.0898 & 0.3803 \\
0.8       & 0.6063 & 0.0865 & 0.3971 \\
1.0       & \textbf{0.6021} & \textbf{0.0832} & \textbf{0.4075} \\
\hline\hline
\end{tabular}
\caption{Effect of Proportion of Unremoved Nodes $\alpha$  on MSE, MAE, and R-squared}
\end{table}

The ablation studies confirm that integrating knowledge graph produced by GraphRAG significantly improves startup success prediction. Adjusting regularization strength ($\lambda$) enhances accuracy, while removing key graph edges reduces performance, highlighting the importance of relational data like collaboration and competition. These results validate the effectiveness of our GraphRAG-based approach in handling sparse data and low signal challenges in venture capital analysis.

%% file: tex/conclusion.tex
\section{Limitation}
GraphRAG's performance relies heavily on data quality, with high costs for processing unstructured text and computational challenges during indexing. Large, unfiltered knowledge graphs increase resource consumption, and the model's generalization beyond Chinese A-share datasets needs further validation.

\section{Future Work}
While our proposed GraphRAG-enhanced multivariate time series analysis method has shown promise in predicting startup success, there are several areas that warrant further investigation to address current limitations and enhance the model's capabilities.

\textbf{Generalization Across Datasets:} A key limitation of our current work is the reliance on non-structured data, which can be difficult to obtain consistently across different regions and industries. The dataset used in this study is primarily focused on the Chinese A-share market, which may limit the generalizability of our findings. Future research should focus on testing the model’s performance on datasets from other countries and markets to evaluate its robustness and adaptability across different economic and cultural environments.

\textbf{Exploration of Soft Masking:} The current approach utilizes a specific method for constructing the mask matrix based on the Leiden algorithm. However, there is potential to explore other soft masking techniques that might offer improved performance. Future work could investigate the use of different clustering algorithms to generate mask matrices, potentially enhancing the model’s ability to capture complex inter-company relationships.

\section{Conclusion}
This paper proposed a GraphRAG augmented Seq2Seq multivariate time series method to predict startup success in venture capital. By integrating GraphRAG with traditional time series models, we tackled a key issue often overlooked in previous studies: the inter-company relationships. Experimental results show significant improvements over existing methods, with GraphRAG enhancing reasoning by leveraging knowledge graphs and retrieval to capture competitive and cooperative dynamics, improving accuracy and robustness under sparse data conditions. Our approach not only mitigates some challenges related to data sparsity but also provides more accurate decision support for venture capitalists.